\documentclass[runningheads]{llncs}
\usepackage{graphicx,hyperref}

\begin{document}

\title{Another virtue of wavelet forests?}

\author{Christina Boucher~\inst{1} \and
Travis Gagie~\inst{2} \and
Aaron Hong~\inst{1} \and\\
Yansong Li~\inst{2} \and
Norbert Zeh~\inst{2}}

\authorrunning{C.\ Boucher et al.}

\institute{Department of Computer and Information Science and Engineering, \\University of Florida, Gainesville, USA \and
Faculty of Computer Science,  Dalhousie University, Halifax, NS, Canada}

\maketitle

\begin{abstract}
A wavelet forest for a text $T [1..n]$ over an alphabet $\sigma$ takes $n H_0 (T) + o (n \log \sigma)$ bits of space and supports access and rank on $T$ in $O (\log \sigma)$ time.  K\"arkk\"ainen and Puglisi (2011) implicitly introduced wavelet forests and showed that when $T$ is the Burrows-Wheeler Transform (BWT) of a string $S$, then a wavelet forest for $T$ occupies space bounded in terms of higher-order empirical entropies of $S$ even when the forest is implemented with uncompressed bitvectors.  In this paper we show experimentally that wavelet forests also have better access locality than wavelet trees and are thus interesting even when higher-order compression is not effective on $S$, or when $T$ is not a BWT at all.

\end{abstract}

\section{Introduction}
\label{sec:introduction}

A wavelet tree~\cite{GGV03,Nav14} for a text $T [1..n]$ over an alphabet $\sigma$ stores $T$ in $n H_0 (T) + o (n \log \sigma)$ bits and supports access, rank and select on $T$ in $O (\log \sigma)$ time, where $H_k (T)$ is the $k$th-order empirical entropy of $T$, $T.\mathrm{rank}_c (i)$ is the number of copies of $c$ in $T [1..i]$ and $T.\mathrm{select}_c (i)$ is the position of the $i$th copy of $c$ in $T$.  One important application of wavelet trees is in FM-indexes~\cite{FM05}, for which we use rank queries over the Burrows-Wheeler Transform (BWT)~\cite{BW94} of the indexed string.

Ferragina et al.\cite{FGMS05} showed that when $T$ is the BWT of a string $S$ and we carefully split $T$ into blocks of varying lengths, then storing a wavelet tree for each block takes a total of $n H_k (S) + o (n \log \sigma)$ bits for $k = o(\log_\sigma n)$.  M\"akinen and Navarro~\cite{MN07} then showed that a single wavelet tree for $T$ implemented with RRR-compressed bitvectors~\cite{RRR07} achieves the same space bound.  Finally, K\"arkk\"ainen and Puglisi~\cite{KP11} showed that splitting $T$ into the right number of fixed-length blocks and storing a wavelet tree for each block together with the rank of each distinct character at the beginning of each block, achieves the same space bound even when the wavelet trees are implemented with uncompressed bitvectors.  Uncompressed bitvectors are faster than compressed ones, so K\"arkk\"ainen and Puglisi's scheme is used in some of the most competitive implementations of FM-indexes~\cite{GKKPP19}.

K\"arkk\"ainen and Puglisi called their scheme {\em fixed-block compression boosting} but, because we think it may be useful even in cases when higher-order compression is not effective, we use the name {\em wavelet forest}.  One such case is storing a single human genome for DNA alignment, which is the main practical application of FM-indexes; for example, Bowtie~\cite{LTPS09,LS12} and BWA~\cite{LD09} store their BWTs in $\lg \sigma = 2$ bits per character.  Can wavelet forests help even here?

One of the main weaknesses of FM-indexes is their poor access locality~\cite{CHSTV15}: searching for a pattern $P [1..m]$ involves a sequence of $\Theta (m)$ rank queries at positions usually scattered throughout the BWT, and these queries must be performed in the right order.  We might expect a search for a pattern $P [1..m]$ to cause $\Theta (m)$ cache misses when $n$ is large, but the situation can be even worse.  If $H_0 (T) \approx \lg \sigma$ then a wavelet tree for $T$ supports an average rank query on $T$ as rank queries on about $\lg \sigma$ bitvectors, each of length $n$ --- so searching for $P$ can conceivably cause $\Theta (m \log \sigma)$ cache misses on average.

\begin{figure}[t]
\begin{center}
\includegraphics[width=.7\textwidth]{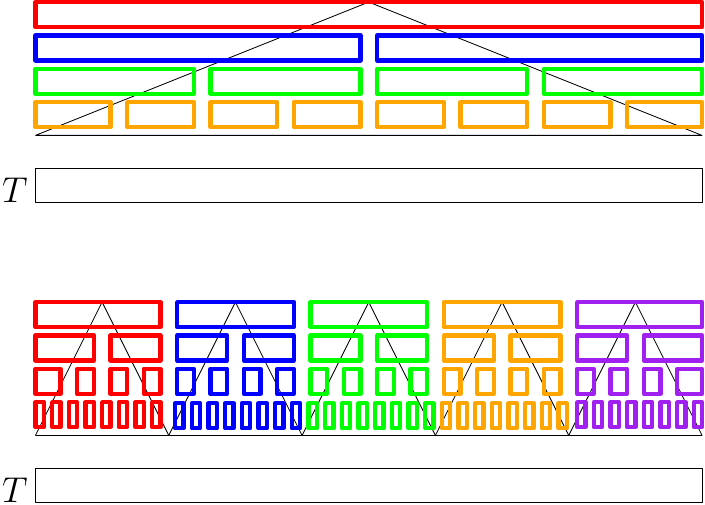}
\caption{A wavelet tree {\bf (top)} and a wavelet forest {\bf (bottom)} for $T$ over an alphabet of size 16.  Coloured blocks indicate bitvectors.  If $T$ is between 4 and 5 times too big for cache, then querying bitvectors of different colours will usually cause a cache miss, but querying bitvectors of the same colour (which are stored together) should not.  A root-to-leaf descent to answer an access or rank query will usually cause 4 cache misses in the wavelet tree but (perhaps) only one in the wavelet forest.}
\label{fig:tree_vs_forest}
\end{center}
\end{figure}

Suppose we store a wavelet forest for $T$ with block length $b$.  Then a rank query on $T$ is again supported as rank queries on about $\lg \sigma$ bitvectors, but with standard implementations all the bitvectors in the wavelet tree for a block are stored together in $O (b \log \sigma)$ bits.  As long as $b$ is not too big, we conjecture that searching for $P$ may now actually cause only $\Theta (m)$ cache misses.  Our idea is illustrated in Figure~\ref{fig:tree_vs_forest}.  If this really works then access should speed up as well, as it is supported similarly and does not even need ranks stored at the beginning of each block.  In short, we think wavelet forests are likely interesting even when higher-order compression is ineffective on $S$, or when $T$ is not a BWT at all.

\section{Experiments}
\label{sec:experiments}

We built several wavelet forests and measured how quickly they supported access compared to single wavelet trees.  We used incompressible data, so any speedup cannot be due to compression boosting.

First, we pseudo-randomly generated an 8 GB dataset, which we interpreted variouslyl as about 64 billion bits, 32 billion base-4 digits, 24 billion octal digits, 16 billion hexadecimal digits and 8 billion ASCII characters.  For each interpretation of the dataset, we used the {\tt sdsl-lite} library~\cite{gbmp2014sea} to build a single wavelet tree (with uncompressed bitvectors) over the dataset, and to build several forests of wavelet trees (also with uncompressed bitvectors) over the dataset.

We used Huffman-shaped wavelet trees because SDSL's balanced wavelet trees seem to have height 8 regardless of the alphabet size, whereas Huffman-shaped wavelet trees will have expected depth at most about $\lg \sigma + 1$, where $\sigma$ is the effective alphabet size.  For each interpretation of the dataset, we built forests with 0.1 MB blocks, 1 MB blocks and 10 MB blocks; interpreting the dataset as ASCII characters --- the largest alphabet --- we also built forests with 0.05 MB blocks, 0.5 MB blocks and 5 MB blocks.

We built the wavelet trees and wavelet forests with {\tt g++} version 12.2.0 on a server with an AMD EPYC 75F3 CPU running Red Hat Enterprise Linux 7.7, then saved them to file.  To ensure noticeable cachine effects, however, we performed our actual experiments on a virtual machine allocated 4 GB of RAM and 100 GB of swap space on a laptop with an Intel Core i7-11700 CPU, both running Ubuntu 22.04.2 Linux.  Our source code is available at \url{https://github.com/AaronHong1024/wavelet_forest}.

\begin{figure}[t]
\begin{center}
\includegraphics[width=.7\textwidth]{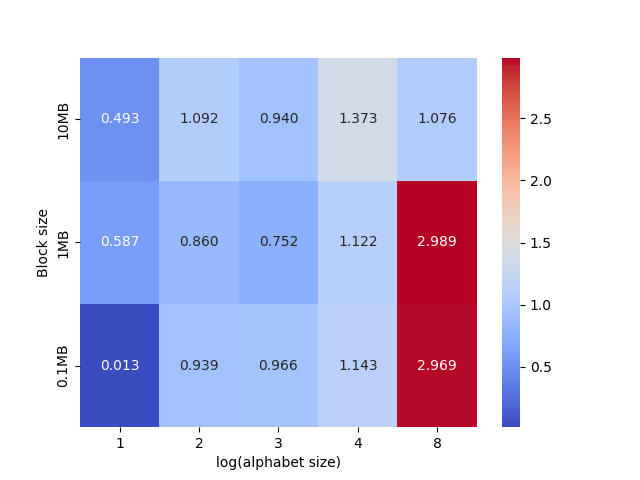}
\caption{The ratio of the times taken by wavelet trees and wavelet forests with 0.1 MB, 1 MB and 10 MB blocks, to access 10\,,000 characters (in the appropriate alphabet).}
\label{fig:heatmap}
\end{center}
\end{figure}

For each wavelet tree and forest, we pseudo-randomly chose 10\,,000 characters (in the appropriate alphabet) from the dataset and measured the total time to access them.  Our results for forests with 0.1 MB blocks, 1 MB blocks and 10 MB blocks are shown in Figure~\ref{fig:heatmap}.  Wavelet forests were slightly faster --- presumably because of better access locality --- for hexadecimal for all those block lengths; they were also very slightly faster for ASCII with 10 MB blocks, and about 3 times faster with 0.1 MB blocks and 1 MB blocks.

\begin{figure}[t]
\begin{center}
\includegraphics[width=0.7\textwidth]{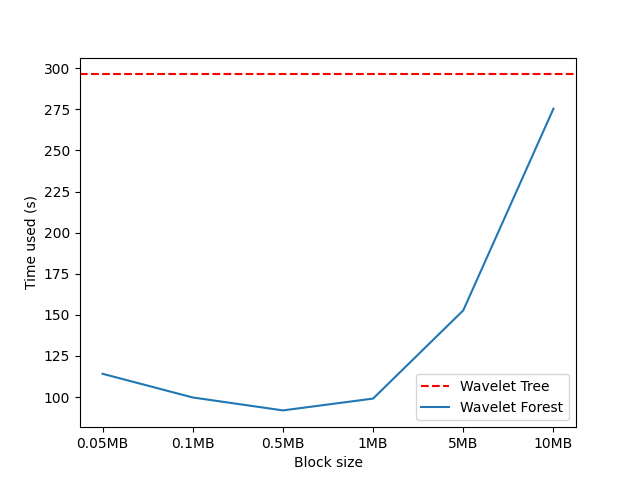}
\caption{The total time for accessing 10\,,000 characters in an 8 GB ASCII dataset with a wavelet tree and wavelet forests, with block lengths ranging from 0.05 MB to 10 MB.}
\label{fig:linear}
\end{center}
\end{figure}

As the largest alphabet, we examined the access times for ASCII using more block sizes, with the results shown in Figure~\ref{fig:linear}.  The access time for the wavelet tree does not depend on the block length, so we display it as a horizontal line.  Interestingly, although we expected access locality to improve monotonically as the block length decreased, the access time was least for the wavelet forest with 0.5 MB blocks (noticeably less than with 0.05 MB blocks).

\section{Future work}
\label{sec:future_work}

Our experiments seem to show that, as we conjectured, wavelet forests have better access locality than wavelet trees, particularly for larger alphabets.  We are surprised, however, that Figure~\ref{fig:linear} seems to show that the best speedups are obtained with fairly large block lengths.  We leave as future work investigating why this is the case.  In the meantime, we plan to repeat our experiments with rank queries --- which require storing the ranks for all the characters at the beginning of each block --- in addition to access queries, and report the results in the full version of this paper.  

We started investigating the access locality of wavelet forests after a conversation with Jouni Sir\'en led us to conjecture that reordering the blocks of a BWT into clusters could reduce the number of long-distance LF steps.  To test this idea, we divided the BWT of the 15 GB Douglas Fir genome into blocks and clustered them with Metis~\cite{KK98} and found that, for reasonable block and cluster sizes, we could keep most of the LF steps within the same clusters.

We built a virtual machine with 1 GB of RAM and tried extracting substrings from an FM-index for the Douglas Fir genome, both with a standard BWT and with the BWT blocked and clustered.  We achieved a significant speedup in practice --- even when the block and cluster sizes were the same and we did not cluster the blocks.  We eventually realized that blocking by itself was giving us the speedups, with clustering working in theory but not in practice.  While this realization led to this paper, we would still like to know if clustering can have a practical benefit.

\section*{Acknowledgments}

Many thanks to Simon Gog, Ben Langmead and Jouni Sir\'en for helpful conversations.

\end{document}